\documentclass{article}

\usepackage{amsmath}
\usepackage{amsfonts}
\usepackage{amssymb}

\begin{document}

\title{Moving Charges, Detectors and Mirrors in a Quantum Field with
Backreaction}
\author{Chad R. Galley\footnote{crgalley@physics.umd.edu}, B. L. Hu\footnote{hub@physics.umd.edu}, and Philip R. Johnson\footnote{philipj@physics.umd.edu}}
\date{}

\maketitle

{\it Department of Physics,
University of Maryland,
College Park, MD 20742, USA. 

-- Invited paper presented by BLH at the Third International
Symposium on Quantum Field Theory under the influence of External
Conditions, Oklahoma City, Sept. 2003. Proceedings edited by
Kimball Milton (Rinton Press, 2004).
}

\bigskip

%\maketitle

\begin{abstract}
This is a progress report on our current work on moving
charges\cite{JH1,JH2}, detectors\cite{RHA}, and moving
mirrors\cite{RHK,GH1} in a quantum field treated in a fully
relativistic way via the Feynman-Vernon influence functional
method\cite{InfluenceFunctional}, which preserves maximal quantum
coherence of the system with self-consistent back-reaction from
the field.
\end{abstract}

%\maketitle

\section{Introduction}

The interplay of moving charges, detectors or boundaries
(mirrors) with a quantum field (scalar and electromagnetic
considered) can be analyzed at different levels of precision and
sophistication.   At the lowest level of approximation described
in textbooks, one computes the charge motion as determined by a
fixed background field. One can then examine the modifications in
the field by the moving charge, which take the forms of emitted
classical radiation, acceleration field, polarization cloud and
changing field correlations. A detailed account of how these
familiar classical behavior arise from a field of quantum origin
is already a non-trivial task. For a self-consistent treatment
one must allow for full backreaction of the quantum field on the
charge. At the classical level, a well-known backreaction effect
is radiation reaction. At the quantum level additional
backreaction effects arise from quantum fluctuations in the form
of quantum dissipation. (Claims that classical radiation reaction
can be derived from quantum fluctuations are
misleading.\cite{CAPRI}) The importance of a coherent
self-consistent  treatment of the mutual interaction between a
particle and a quantum field arises in many situations, becoming
particularly important under relativistic and strong field
conditions.(For QED see, e.g., \cite{KME})

A deep result of Unruh\cite{Unr76} in 1976 states that a
uniformly accelerated detector (a physical object with some
internal degree of freedom which can respond to external
influences, such as a two-level atom) registers thermal radiance
at a (Unruh) temperature $T_U$ proportional to its proper
acceleration. This can be interpreted as a kinematical effect,
i.e.,  amplification of quantum noise (vacuum fluctuations of the
quantum field) by the moving detector. Its exact thermality
arises from an exponential red-shifting brought about by the
uniform acceleration\cite{Dalian}. The cases of nonthermal
radiation in non-uniformly accelerated detectors have also been
studied\cite{RHA,RHK}.

It was known even earlier that imposition of boundary conditions
on a quantum field such as introducing two conducting plates
(mirrors) or dielectric surfaces also leads to discernible
physical effects. A celebrated example is the attractive Casimir
force between two parallel conducting plates which is discussed
extensively in this conference series. The effects related to a
moving mirror\cite{FulDav} is sometimes referred to as the
dynamical Casimir effect, which has close similiarity with
cosmological particle creation. A moving mirror imparts a
changing boundary condition on the quantum field which has
detectable effects nearby and afar.

Investigations of these three cases: moving charges, moving
detectors, and moving mirrors carry both theoretical and practical
values. In addition to the issues of radiation reaction and vacuum
fluctuations mentioned above, these problems probe directly into
the physical nature of the vacuum, such as vacuum
viscosity\cite{Zel70} in relation to the backreaction of
cosmological particle creation or vacuum friction\cite{Kardar} in
the dynamical Casimir effect. Moving detectors and moving mirrors
were studied in the 70's as analogs to the Hawking effect from
black holes.\cite{Haw75,FulDav,Unr76} Readers interested in
coherent back-action of a cavity quantum field on a moving atom
are referred to a companion paper\cite{HRS} dealing with the
Casimir-Polder effect\cite{CasPol} based on moving atoms through
a quantum field in the presence of a mirror\cite{SHCasPol} and an
analysis of a recent proposal for the detection of the Unruh
effect in a QED cavity.\cite{Scully}

\section{Self-Consistent Trajectories of Moving Charges in a Quantum Field}

The problem of self-consistent radiation reaction between particle and field
is, of course, an old one. The famous Abraham-Lorentz-Dirac (ALD) equation
obtained for classical theory is ultimately inconsistent at the point
particle level, giving either runaway or pre-accelerating (acausal) solutions%
\cite{ClassicalALDRR}. Markovian quantum treatments for nonrelativistic
particles have given essentially the same results\cite{QuantumALDRR}: a
charge with structure, or a UV regulated field, can yield consistent
results, but pathologies return in the point particle limit. Below, we
discuss the connections between causal and pathology-free particle motion,
self-consistent non-Markovian quantum backreaction, and the stochastic
regime for backreaction.

\subsection{Relativistic worldlines and quantum fields}

Consider a spinless particle of charge $e$ and mass $m_{0}$ moving through
a massless scalar field $\phi $ with action
\begin{equation}
S_{\phi }+S_{int}+S_{x}=\frac{1}{2}\int dy\left( \partial _{\mu }\phi
\right) ^{2}+\int dyj\left( y\right) \phi (y)+\int d\tau u\left( \tau
\right) \left[ m_{0}+V\left( x\right) \right] ,
\end{equation}%
where the charge current $j(y,x]=e\int d\tau u(\tau )\delta \left( y-x\left(
\tau \right) \right) $ is a functional of the particle's parametrized
spacetime trajectory $x^{\mu }\left( \tau \right) ,$ $V\left( x\right) $ is
an external field, and $u\left( \tau \right) =\left( \dot{x}^{\mu }\dot{x}%
_{\mu }\right) ^{1/2}$ is the reparametrization-invariant factor giving the
equations of motion for a relativistic particle. The matrix elements of the
quantum evolution operator for the particle-field system may be obtained by
summing over all field and particle (i.e. worldline) histories with
amplitudes determined by the above action.

For particle-motion backreaction problems, the worldline path integral
representation can be more efficient than the usual quantum field theoretic
techniques because it directly employs only the particle's trajectory rather
than the infinite set of field degrees of freedom. This approach, dating
back to Feynman and Schwinger,\cite{Worldline: Feynman and Schwinger} has
been applied to a range of problems from action at a distance QED,\cite%
{Action at a distance QED} particle production and high-order
background field QED and QCD calculations,\cite{Worldline in
background field QED and QCD} to studies of finite size
effect\cite{SOH}, reparametrization invariance in quantum
cosmology\cite{Hartle and Kuchar} and string
theory\cite{StringTheory}. We use the worldline path integral
representation together with the Feynman-Vernon influence
functional formalism for describing particle motion with
self-consistent backreaction. We employ an open-systems approach
by coarse-graining the field to obtain the influence functional,
from which we find equations of motion for the particle
trajectory.

\subsection{Coarse-grained effective action for backreaction on worldlines}

For technical convenience we make the simplifying assumption that at an initial
time $t_{i}$ the states of the particle and field factorize: $\hat{\rho}%
\left( t_{i}\right) =\hat{\rho}_{x,i}\otimes \hat{\rho}_{\phi ,i}.$ We also
introduce a UV\ momentum cutoff $\Lambda $ for the field.\footnote{%
While more general initial conditions are possible, no fully satisfactory
method has yet been developed to describe completely physical initial
correlated states for the kind of particle-field systems considered here\cite%
{RomeroPaz96}.}

The full particle-field system evolves unitarily according to 
\begin{eqnarray}
	\hat{\rho} \left( t_{f}\right) =\hat{U}\left( t\right) \left( \hat{\rho}_{x,i}\otimes
\hat{\rho}_{\phi ,i}\right) \hat{U}\left( t\right) ^{\dagger } . \nonumber
\end{eqnarray}
 The reduced
density matrix of the particle at time $t_{f}$ is found by integrating out
the field degrees of freedom: $\hat{\rho}_{x}\left( t_{f}\right) =Tr_{\phi }%
\hat{\rho}\left( t_{f}\right) .$ It can be expressed as
\begin{equation}
\rho _{x}\left( x_{f},x_{f}^{\prime };t_{f}\right) =\int
dx_{i}dx_{i}^{\prime }J_{r}\left( x_{f},x_{f}^{\prime };x_{i},x_{i}^{\prime
}\right) \rho _{x}\left( x_{i},x_{i}^{\prime };t_{i}\right) ,
\end{equation}%
where $J_{r}$ is the density matrix evolution operator. It has a path
integral representation of the form%
\begin{equation}
J_{r}\left( x_{f},x_{f}^{\prime };x_{i},x_{i}^{\prime }\right)
=\int_{x_{i},x_{i}^{\prime }}^{x_{f},x_{f}^{\prime }}DxDx^{\prime }\mathcal{M%
}\left[ x,x^{\prime }\right] e^{\frac{i}{\hbar }S_{CGEA}\left[ x,x^{\prime
}\right] },  \label{Reduced density matrix evolution kernel}
\end{equation}%
where the path integrals involve summing over worldlines histories, with the
appropriate boundary conditions and the class of allowed paths enforced by
the functional measure $\mathcal{M}\left[ x,x^{\prime }\right] .$ The
coarse-grained effective action\cite{cgea} for the particle degrees of
freedom is
\begin{equation}
S_{CGEA}=S\left[ x\right] -S\left[ x^{\prime }\right] +S_{IF}\left[
x,x^{\prime }\right] ,  \label{Coarse-grained effective action}
\end{equation}%
the influence action $S_{IF}$ is defined by
\begin{equation}
S_{IF}\left[ x,x^{\prime }\right] =-i\hbar \ln \mathcal{F}\left[ x,x^{\prime
}\right] ,
\end{equation}%
and $\mathcal{F}\left[ x,x^{\prime }\right] $ is the Feynman-Vernon
influence functional. Since determining the influence action $S_{IF}$ only
requires integrating-out the field path integrals, we will not need the
explicit form of $\mathcal{M}$ here. For Gaussian field states, $S_{IF}$ is
given exactly by%
\begin{equation}
S_{IF}=-\int \int dydy^{\prime }\left[ j^{-}(y)G_{R}(y,y^{\prime
})j^{+}(y^{\prime })+j^{-}(y)G_{H}(y^{\prime },y^{\prime })j^{-}(y^{\prime })%
\right] ,
\end{equation}%
where $G_{R,H}$ are the retarded and Hadamard Green's functions of the
scalar field, respectively, and $j^{\pm }\left( y\right) =j(y,x]\pm
j(y,x^{\prime }]$.

We consider first the semiclassical limit, described by a pathalogy-free
time-dependent modification of the ALD equation, and then discuss the
ALD-Langevin equations describing the quantum field-induced fluctuations in
the motion of the particle around its average (semiclassical) trajectory.

\subsection{Semiclassical level: ALD equation and non-Markovian cure of its
pathologies}

If the initial particle state is well-localized in phase-space, then 
\begin{eqnarray}
	\left( \delta S_{CGEA}/\delta x^{-}\right) |_{x^{\pm }=0}=0 \nonumber
\end{eqnarray}
gives the following
semiclassical equations of motion for the average trajectory\cite{JH1}:%
\begin{equation}
m\left( \tau \right) \ddot{x}_{\mu }\left( \tau \right) -\partial _{\mu
}V(z)=f_{\mu }^{R.R.}\left( \tau \right) =e^{2}g\left( \tau \right) \left(
\dot{x}_{\mu }\ddot{x}^{2}+\dddot{x}_{\mu }\right) +\mathcal{O}\left(
\Lambda ^{-1}\right) ,  \label{ALD equation}
\end{equation}%
where the $\mathcal{O}\left( \Lambda ^{-1}\right) $ represents
higher-derivative terms which are suppressed at low energies. The worldline
parameter $\tau $ is chosen, for simplicity, so that $\tau \left(
t_{i}\right) =0.$ Equation (\ref{ALD equation}) has the usual ALD form
except for the coefficients $m\left( \tau \right) $ and $g\left( \tau
\right) $ whose time-dependence is a consequence of dynamical redressing of
the particle by the field on a time-scale determined by the cutoff $\Lambda$.
This redressing is a consequence of our having assumed an initially
factorized particle-field state. For late times defined by $\tau \gg m_{0}r_{0}/\Lambda $,
the particle is dressed, the dynamics are effectively those of a physical particle state
with $g\left( \infty \right)
=1$ and renormalized mass $m(\infty )=m(0)-\kappa e^{2}\Lambda /8\pi ;$ $%
\kappa $ is a constant of order one that depends on the details of how the
UV field is regulated.\cite{JH1} For the solutions to be runaway free at
late times the bare mass must satisfy $m\left( 0\right) =m_{0}>\kappa
e^{2}\Lambda /8\pi $ (e.g. see the analysis of causality by Ford, Lewis, and
O'Connell\cite{FOL}).

The short-time regime is of interest because this is where acausality of the
traditional ALD\ equation arises. Taking into account the dynamical
redressing of the particle, we find $g\left( 0\right) =0,$ and thus $f_{\mu
}^{R.R.}\left( 0\right) =0.$ The suppressed $\mathcal{O}\left( \Lambda
^{-1}\right) $ terms are also time-dependent, but they too have the generic
property of vanishing at $\tau =0$. Consequently, Eq. (\ref{ALD equation})
has a causal solution that is uniquely determined by initial position, $%
x\left( 0\right) ,$ and velocity, $\dot{x}\left( 0\right) ,$ data. The time
scale for radiation reaction forces to grow from zero to their asymptotic
ALD value is given by $m_{0}r_{0}/\Lambda ,$ which is bounded from below by
the time it takes for light to cross the classical radius of the particle.

\subsection{Stochastic Level: Decoherence and the ALD-Langevin equation}

When decoherence of the quantum histories is sufficiently strong, the
worldline path integrals may be approximated by replacing the coarse-grained
effective action with a stochastic effective action $S_{\chi }.$ Then $%
\left( \delta S_{\chi }/\delta z^{-}\right) |_{z^{\pm }=0}=0$ gives
ALD-Langevin (ALDL) equations of motion, where $z\left( \tau \right) $ is
the fluctuation (deviation coordinate) of the particle around the average
worldline $x\left( \tau \right) .$

To find a self-consistent ALDL equation we first solve for the
quantum-average worldline $x\left( \tau\right) ,$ including the effects of
radiation reaction, by using Eq. (\ref{ALD equation}). The ALDL equations
are then%
\begin{equation}
\eta\left( \tau\right) =m\ddot{z}_{\mu}\left( \tau\right) +z_{\nu}\frac{%
\partial V\left( x\right) }{\partial x^{\mu}\partial x_{\nu}}-\frac{e^{2}}{%
8\pi}\left( S_{\mu\nu}\dot{z}^{\nu}+R_{\mu\nu}\dddot{z}^{\nu }\right) ,
\label{ALDL equation}
\end{equation}
where $\eta\left( \tau\right) $ is the stochastic noise. At late-times, $%
R_{\mu\nu}=\left( g_{\mu\nu}-\dot{x}_{\mu}\dot{x}_{\nu}\right) $ and $%
S_{\mu\nu}=\left( \ddot{x}^{2}g_{\mu\nu}-\dot{x}_{\mu}\dddot{x}_{\nu
}\right) .$ At $\tau=0,$ $R$ and $S$ vanish and the stochastic equations are
causal and unique. The noise is given by
\begin{equation}
\eta_{\mu}\left( \tau\right) =e\left( \ddot{x}_{\mu}\left( \tau\right) +\dot{%
x}^{\nu}\dot{x}_{[\nu}\partial_{\mu]}\right) \chi\left( x\left( \tau\right)
\right) ,  \label{single particle noise}
\end{equation}
where $\chi\left( y\right) $ is a stochastic field evaluated along the
average worldline, whose noise correlator is
\begin{equation}
\left\langle \left\{ \chi(y),\chi\left( y^{\prime}\right) \right\}
\right\rangle =\hbar G^{H}\left( y,y^{\prime}\right) .
\end{equation}
Note that the stochastic field correlations are nonlocal since the Hadamard
function is non-vanishing for spacelike separated points. Not surprisingly,
quantum fluctuation induced noise is colored (nonlocal).

\subsection{Radiation reaction and quantum fluctuations}

At the semiclassical level described by the ALD equation there is radiation
reaction independent of quantum fluctuations [i.e. there is no $\hbar $ in
Eq. (\ref{ALD equation})]. At the stochastic level described by the
ALD-Langiven (ALDL) equation, there are additional quantum backreaction
effects manifesting as quantum dissipation which is balanced by quantum
fluctuations. Historically there are incorrect claims that (classical)
radiation reaction is balanced by quantum fluctuations. This mistake comes
from the failure to distinguish between these two levels of backreaction
effects.\cite{CAPRI}

As an explicit example illustrating the difference between semiclassical and
stochastic radiation reaction, consider an external field that uniformly
accelerates a charged particle through the scalar field vacuum. It is a
well-known consequence of the semiclassical ALD equation that the (average)
radiation reaction force vanishes, even though the particle does radiate.
However, the ALDL equation, describing the quantum field-induced
fluctuations in the particle motion around it's average worldline, shows
that the particle responds to the scalar vacuum as if it were a thermal
state at the Unruh temperature.\cite{Unr76} These quantum fluctuations are
balanced by the (non-vanishing) dissipative backreaction term in the ALDL
equations, despite the fact that the semiclassical radiation reaction force
is zero.

Analysis of the \textit{nonrelativistic} limit for QED has shown
that Bremsstrahlung is the primary source of decoherence for
charged particles. An ALD-Langevin equation for
\textit{nonrelativistic} QED in the late time limit has also been
derived by a number of authors.\cite{FOL} Currently PRJ and BLH
are in the process of extending these results and methods to a
relativistic particle in the quantum electromagnetic
field.\cite{JH2} This
formulation should provide a way for describing \textit{%
relativistic} particle backreaction, including quantum-induced
stochastic fluctuations, under conditions of sufficient
decoherence.

\section{Moving mirror and detector in a quantum field with backreaction}

We now turn our attention to moving mirrors in a quantum field. As
background material refer to\cite{FulDav,RHA,RHK}. Consider in $d+1$
dimensional flat spacetime a point-like detector moving through
and (linearly) coupled to a massive scalar field that is
constrained to vanish at the $n$-dimensional
surface of a mirror ($n < d$). To allow for backreaction from the field, the detector and
mirror are assumed to move along arbitrary, unprescribed
trajectories determined by the dynamics under the influence of
the quantum field in a self-consistent manner. In the open system approach
the influence functional, which captures the effects of the coarse-grained quantum field,
generates correlation functions of the {\it constrained} field at
various times along the detector worldline and the variation of
the related stochastic effective action leads to the stochastic equations of motion
for both objects and their responses.

Denote by $x_1 ^\alpha (\tau_1)$ and $x_2 ^\beta (\tau_2, {\vec \sigma}_2 )$ the detector
and mirror paths with affine parameters $\tau_1, \tau_2$, respectively, with the shape of
the mirror parametrized by ${\vec \sigma}_2$, by $Q_1$ the internal degree of
freedom of the detector assumed to be a harmonic oscillator with
natural frequency $\Omega$, and by $j_1[Q_1 (\tau_1) ]$ the
detector portion of the interaction with the constrained field
$\phi_c (x)$. The total action describing the motions of the
detector and the mirror, the excitation of the detector (by an
energy $\hbar \Omega$), and the quantum field constrained by the
moving mirror, is given by\footnote{Natural units are used so that $\hbar = c = 1$
and the flat spacetime metric is $\eta _{\mu \nu} = {\rm diag}(1, -1, \ldots, -1)$.}
\begin{eqnarray}
   S_{tot} &=& S_{\phi_c} [\phi_c] + S_Q [Q_1] + S_{x_1} [x_1] + S_{x_2} [x_2]
   + S_{int} [ j_1 [Q_1] , x_1, x_2, \phi_c] \nonumber \\
        &=& \frac{1}{2} \: \int d^{d+1} y \: \left\{ \partial_\mu \phi_c (y) \:
        \partial ^\mu \phi_c (y) -m^2 \phi_c ^2 (y) \right\} \nonumber \\
       &&+ \frac{1}{2} \: \int d\tau_1 \left\{ {\dot Q}_1 ^2 (\tau_1) \: N_1 ^{-1} (\tau_1)
       - \Omega^2 Q_1 ^2 (\tau_1) \: N _1 (\tau_1) \right\} \nonumber \\
       &&- \frac{m_1}{2} \: \int d\tau_1\: \left\{ {\dot x}_1 ^\mu (\tau_1) \:
       {\dot x}_{1 \mu} (\tau_1) \: N_1 ^{-1} (\tau_1) + N_1 (\tau_1) \right\} \nonumber \\
    &&+ S_{x_2}[x_2] + e \int d\tau_1 \: j_1 [Q_1 (\tau_1) ] \:
\phi_c ( x_1 ^\alpha (\tau_1) ) ~. \label{action}
\end{eqnarray}
The lapse function along the detector worldline is $N_1 (\tau_1)$ and is closely
related to the induced metric along the worldline $h_1 (\tau_1) = {\dot x}_1 ^\mu \:
{\dot x}_{1 \mu}$ with an overdot denoting differentiation with respect to $\tau_1$.
The factor $j_1 [ Q_1 (\tau_1) ] \: d\tau_1$ is a
reparametrization-invariant coupling that describes various
interactions with the constrained field, e.g. the monopole
coupling $j_1 = \sqrt{h_1} \: Q_1$ and the minimal coupling $j_1
= {\dot Q}_1$. The mirror action $S_{x_2}[x_2]$ is difficult to write
down for the general case of an $n$-dimensional object moving relativistically
in which case the mirror cannot be regarded as infinitely rigid. In any case,
the explicit form of $S_{x_2}$ is irrelevant at the formal level of this discussion.
In what follows, a $\phi$ appearing without a subscript $_c$ denotes a field
configuration that does not respect the constraint imposed by the mirror.

Note that there is no explicit interaction Lagrangian appearing
in $S_{tot}$ describing the {\it direct} coupling of the mirror to
the field. Rather, it is through the Dirichlet boundary conditions
in the constrained field configurations in (\ref{action}). The
interaction term $e \int d\tau_1 \: j_1 [Q_1 (\tau_1) ] \: \phi_c
( x_1 ^\alpha (\tau_1) )$  describes the mutual effects of the
mirror motion on the field as $\phi_c$ couples to the detector
through $j_1$ and the detector's motion. While this is linear in
$\phi_c$ it is highly nonlinear as a function of the detector
worldline $x_1$. The manifestation of the mirror worldvolume
$x_2$ appears here in that $j_1$ couples to the {\it constrained}
field which is evaluated at the position of the detector ${\vec
x}_1$. The constrained field transmits the information and
mediates the influence on the motion of the mirror and the
detector together. These effects must be determined
self-consistently which is why the influence functional formalism
is so valuable in studying problems like this one involving
backreaction dynamics.

In the open system approach the influence functional ${\cal F}_c$
describes the coarse-grained effect of  the constrained field
${\cal E}= (\phi_c)$,  regarded as the `environment',  on the
self-consistent evolution of the detector internal degree of
freedom and the paths of the detector and mirror, taken together
as the `system'  ${\cal S} = (Q_1, x_1,x_2)$. In the path integral
representation it is given by
\begin{eqnarray}
&& {\cal F}_c [ j_1, j'_1; x_1 ^\alpha, x_1^{\prime \beta} ; x_2^\gamma , x_2^{\prime \delta} ] \nonumber \\
&& ~~ ~~ = {\rm Tr}_{\phi_c ^f} \left\{ \hat{U}_c ( t_f, t_i;
j_1, x_1, x_2 ] \:
\hat{\rho} _{\phi_c, i} \: \hat{U} _c ^\dagger ( t_f, t'_i; j'_1, x'_1, x'_2]  \right\} \nonumber \\
&& ~~ ~~ = \int _{\Sigma_f} {\cal D} \phi_c ^f ({\vec y}) \int _{\Sigma_i}
{\cal D} \phi_c ^i ({\vec y}) \int _{\Sigma'_i} {\cal D} \phi_c ^{\prime i} ({\vec y}) \:
\rho _{\phi_c, i} ( \phi_c ^i , \phi _c ^{\prime i} ) \nonumber \\
&& {\hskip 0.5in} \times \int _{\phi_c ^i ({\vec y})} ^{\phi_c ^f ({\vec y}) }
{\cal D} \phi (y) \: \delta [ \phi ( x_2 (\tau_2, {\vec \sigma}_2 ) ) ]  \: e^{ \frac{i}{2} \:
\int d^{d+1} y \: \left( \partial_\mu \phi (y) \: \partial ^\mu \phi (y) -m^2 \phi ^2 (y) \right) } \nonumber \\
&& {\hskip 1.5in} \times e^{ i e \int d\tau_1 \: j_1 [Q_1 (\tau_1) ] \:
\phi ( x_1 ^\alpha (\tau_1) ) } \nonumber \\
&& {\hskip 0.5in} \times \int _{\phi_c ^{\prime i} ({\vec y})} ^{\phi_c ^f ({\vec y}) }
{\cal D} \phi'  (y) \: \delta [ \phi ( x ' _2  (\tau_2, {\vec \sigma}_2 ) ) ] \:
e^{ - \frac{i}{2} \: \int d^{d+1} y \: \left( \partial_\mu \phi' (y) \:
\partial ^\mu \phi' (y) -m^2 \phi^{\prime 2} (y) \right) } \nonumber \\
&& {\hskip 1.5in} \times e^{ - i e \int d\tau_1 \: j_1 [Q' _1
(\tau_1) ] \: \phi' ( x_1 ^{\prime \alpha} (\tau_1) ) } \label{inf}
\end{eqnarray}
where the trace is over constrained field configurations at the final time $t_f$.
We assume that at the initial time $t_i$ the combined ${\cal S} +
{\cal E}$ is in a factorized state. Notice that the initial and
final time functional integrals are over constrained field
configurations while the transition amplitudes $U$ are over
unconstrained fields.

The delta functionals constrain the field to
vanish at the surface of the mirror and are introduced by
considering the functional measure ${\cal D} \phi_c (y) = {\cal
D} \phi (y) \: \delta [ \phi ( x_2 ^\alpha (\tau_2, {\vec \sigma}_2 ) )
]$ which does not affect the limits of the integration. A simple
way to realize the delta functional is to introduce an auxiliary
field $Q_2 (\tau_2, {\vec \sigma}_2 )$ on the surface of
the mirror
\begin{eqnarray}
   \delta [ \phi ( x_2 ^\alpha (\tau_2, {\vec \sigma}_2 ) ) ]
   = \int _{-\infty} ^\infty {\cal D} Q_2 \: e^{i \int d\tau_2 \:
   \sqrt{-h_2 (\tau_2, {\vec \sigma}_2)} \: Q_2 ( \tau_2, {\vec \sigma}_2 ) \:
   \phi ( x_2 ^\alpha ( \tau_2 , {\vec \sigma}_2 ) ) } ~.
\label{delta}
\end{eqnarray}
The boundary condition along $x_2$ is therefore implemented as a
Lagrange multiplier on $\phi$ and then summing over all possible
multipliers $Q_2$.

For simplicity we assume the initial state of the field is in a
Gaussian form (e.g. vacuum, thermal). A quick look at (\ref{inf})
using (\ref{delta}) shows that the exponentials are quadratic in
the field. Since the influence functional ${\cal F}_c$ involves
only integrals over the field, it can be evaluated in principle. From
this point onward, there are two possible routes to take: One
can use ${\cal F}_c$ to obtain correlation functions of the {\it
constrained} field at various times along the detector worldline.
Or, by setting to zero the variation of the stochastic effective action,
which is related to the coarse-grained effective action containing ${\cal F}_c$, i.e.,
\begin{eqnarray}
   S_{CGEA} &=& S_{x_1} [x_1] - S_{x_1} [x' _1] + S_{x_2} [x_2] - S_{x_2} [x'_2] \nonumber \\
    && ~~  + S_{Q_1} [Q_1] - S_{Q_1} [Q'_1] -i \: {\rm ln} \: {\cal F}_c ~, \nonumber
\end{eqnarray}
with respect to the appropriate system variable, one obtains a
set of coupled, nonlinear, and nonlocal Langevin equations. This
offers a formal solution to the backreaction problem, but for
arbitrary detector and mirror motions an explicit solution is too
difficult to obtain. It is important, however, to realize that
even though the constrained field cannot be calculated for
arbitrary mirror and detector motions, the influence functional
can be used to construct expectation values of the constrained
field without the knowledge of the form of $\phi_c (y)$.

In many physical situations the mirror mass is sufficiently large
that the recoil experienced by impinging radiation and radiation reaction
is negligible on its center of mass motion. A further
simplification assumes that the detector mass is much larger than
$\hbar \Omega$ so that the recoil effects due to absorption and
emission of quanta do not significantly influence the detector's
motion. In this approximation, the detector and mirror
trajectories have no action and may be prescribed from the
outset. A point-like mirror\footnote{For this particular example
the mirror action describes the free dynamics of a massive point
particle, $S_{x_2} [x_2] = - \frac{m_2}{2} \: \int d\tau_2 \:
\left\{ {\dot x}_2 ^\mu \: {\dot x}_{2 \mu} \: N_2 ^{-1} (\tau_2)
+ N_2 (\tau_2) \right\}$ with $N_2 (\tau_2)$ the mirror lapse
function and differentiation with respect to $\tau_2$ denoted by
an overdot. } moving in one spatial dimension is the case assumed
in most previous work\cite{FulDav,RHK}. One can use this simplest
case to study statistical mechanical features such as the
derivation of generalized fluctuation-dissipation relations and
vacuum friction. Including backreaction effects on the detector
and mirror's (generically relativisitic) motions will allow us to
follow the causal dynamics and the consistent evolution of
correlations and the effect on their trajectories. The
nonrelativistic limit of these results are useful for
quantum/atom optics applications. One area is in quantum computer
designs where maximal quantum coherence of the system in the face
of environmental influences requires a coherent backreaction
treatment. The other area of application is in quantum noise
reduction schemes in interferometer gravitational wave detectors
such as advanced LIGO\cite{BuoChen}.

\section*{Acknowledgments}
BLH thanks Alpan Raval for early discussions on the moving
detector and mirror problem and shows his appreciation to
Mei-Ling Tsang and Jonathan Ozik who made some initial attempts
on the 4-dimensional moving detector and the moving mirror
problems respectively. We thank Albert Roura for discussions.
This work is supported in part by NSF grant PHY03-00710.


\begin{thebibliography}{99}
\bibitem{JH1} Philip R. Johnson and B. L. Hu, Phys. Rev. \textbf{D 65}
(2002) 065015.

\bibitem{JH2} Philip R. Johnson and B. L. Hu, in preparation.

\bibitem{RHA} A. Raval, B.L. Hu, and J. Anglin, Phys. Rev. D53, 7003-7019
(1996).

\bibitem{RHK} A. Raval, B. L. Hu and D. Koks, Phys. Rev. \textbf{D 55}, 4795
(1997).

\bibitem{GH1} C.R. Galley and B. L. Hu,
%\textquotedblleft Moving mirror and detector in a quantum field with coherent backreaction"
in preparation.

\bibitem{InfluenceFunctional} R. Feynman and F. Vernon, Ann. Phys. (N.Y.)
\textbf{24}, 118 (1963).

\bibitem{CAPRI} B.L. Hu and P.R. Johnson, in\textit{\
%Proceedings of the Second Conference on
Quantum Aspects of Beam Physics, }edited by P. Chen
(World Scientific, Singapore, 2001), quant-ph/0012132.

\bibitem{KME}
Y. Kluger, E. Mottola and J. M. Eisenberg, Phys. Rev. D58, 125015
(1998)

\bibitem{Unr76} W.G. Unruh, Phys. Rev. \textbf{D 14}, 870 (1976);
B. S. DeWitt, Phys. Rep., \textbf{19}, 295 (1975);
P. C. W. Davies, J. Phys. \textbf{A}: Gen. Phys. \textbf{8}, 609 (1975);
S. A. Fulling, Phys. Rev. \textbf{D 7}, 2850 (1973).

\bibitem {Dalian}
B. L. Hu, in {\it Proc. 4th Int. Workshop on Thermal Field
Theory} eds. X. Gui and K. Khanna (World Scientific, Singapore,
1996) [gr-qc/9606073]

\bibitem{Zel70} Ya. B. Zeldovich, Zh. Eksp. Teor. Fiz. Pis'ma Red. \textbf{12%
}, 443 (1970) [JETP Lett.\textbf{32}, 307 (1970)]; B. L. Hu, Phys. Lett. \textbf{90A}, 375 (1982).

\bibitem{Kardar} M. Kardar and R. Golestanian, Rev. Mod. Phys. \textbf{71}, 1233 (1999).

\bibitem{Haw75} S.W. Hawking, Commun. Math. Phys. \textbf{43}, 199 (1975).

\bibitem{FulDav} S.A. Fulling, P.C.W. Davies, Proc.\ R. Soc.\ Lond.\ \textbf{%
A 348}, 393 (1976). P. C. W. Davies and S. A. Fulling, Proc. Roy. Soc. Lon.
A356, 237 (1977).

\bibitem{HRS} B. L. Hu, A. Roura and S. Shresta, J. Op. B (2004)

\bibitem{CasPol} H. B. G. Casimir and D. Polder, Phys. Rev. \textbf{73}, 360
(1948).

\bibitem{SHCasPol} Sanjiv Shresta and B. L. Hu, Phys. Rev. A {\bf 68},
062101 (2003).

\bibitem{Scully} M Scully et al, quant-ph/0305178.

\bibitem{ClassicalALDRR} J.D. Jackson, \textit{Classical Electrodynamics }%
(John Wiley and Sons Inc., New York, 1975).

\bibitem{QuantumALDRR} P.W. Milonni, \textit{The Quantum Vacuum }(Academic,
San Diego, 1994).

\bibitem{Worldline: Feynman and Schwinger} R.P. Feynman, Phys. Rev. \textbf{%
80}, 440 (1950); J. Schwinger, \textit{ibid.} \textbf{82}, 664 (1951).

\bibitem{Action at a distance QED} A.O. Barut and I.H. Duru, Phys. Rev.
\textbf{172}, 1 (1989).

\bibitem{Worldline in background field QED and QCD} See, e.g.,  C. Schubert,
hep-th/0101036.

\bibitem{SOH} D.J. O'Conner, C.R. Stephens, and B.L. Hu,
Ann. Phys. (N.Y.) \textbf{190}, 310 (1989).

\bibitem{Hartle and Kuchar} J.B. Hartle and K.V. Kuchar, Phys. Rev. \textbf{%
D 34}, 2323 (1986). J.B. Hartle and K. Schleich, in
\textit{Quantum Field Theory and Quantum Statistics} Vol. 2, eds.
T.A. Batalin etal (Adam Hilgar, Bristol 1987)

\bibitem{StringTheory} M.B. Green, J.H. Schwarz, and E. Witten, \textit{%
Superstring Theory }(Cambridge University Press, Cambridge, 1988).

\bibitem{RomeroPaz96} L. D. Romero and J. P. Paz, Phys. Rev. A \textbf{55},
4070 (1997).

\bibitem{cgea} B. L. Hu, in \textit{Relativity and Gravitation: Classical
and Quantum} Proc. SILARG VII, Cocoyoc, Mexico 1990. eds. J. C. D' Olivo et
al (World Scientific, Singapore 1991).

\bibitem{Yab} E. Yablonovitch, Phys. Rev. Lett. \textbf{62}, 1742 (1989)

%\bibitem{QLE} H. Callen and T. Welton, Phys. Rev. \textbf{83}, 34 (1951); H.
%Mori, Prog. Theor. Phys. \textbf{33}, 423 (1965); G. W. Ford, M. Kac, and P.
%Masur, J. Math. Phys. \textbf{6}, 504 (1965); H. Haken, Rev. Mod. Phys.
%\textbf{80}, 67 (1975); H. Dekker, Phys. Rep. \textbf{80}, 1 (1981).

\bibitem{FOL} G. W. Ford, J. T. Lewis, and R. F. O'Connell, Phys. Rev. Lett.
\textbf{55}, 2273 (1985); Phys. Rev. A \textbf{37,} 4419 (1988).
G. W. Ford and R. F. O'Connell, Phys. Lett. A \textbf{157}, 217
(1991);  Phys. Lett. A \textbf{174}, 182 (1993).

\bibitem{BuoChen}
A. Buonanno and Y. Chen, Phys. Rev. D \textbf{64}, 042006 (2001);
A. Buonanno and Y. Chen, Class. Quantum Grav. \textbf{19}, 1569 (2002).

\end{thebibliography}
\end{document}